\documentstyle [12pt] {article}
\begin{document}
\renewcommand{\baselinestretch}{2.0}
\baselineskip 20pt
\headheight 0.0 in 
\textheight 8.5 in
\centerline{\large{\bf Ground-based detection of a vibration-rotation line
of HD in Orion}}

\vskip 2.0 truecm

\centerline{T. R. Geballe$^{a,*}$, S. K. Ramsay Howat$^{b}$, R.
Timmermann$^{c}$, F. Bertoldi$^{d}$, C. M. Mountain$^{a}$}

\vskip 0.9 truecm

\centerline{\footnotesize $^{a}${\it Gemini Observatory, Hilo, HI 96720,
USA}} 
           
\centerline{\footnotesize$^{b}${\it Astronomical Technology Centre, Royal
Observatory, Blackford Hill, Edinburgh, EH9 3HJ Scotland}}

\centerline{\footnotesize $^{c}${\it Universit\"{a}t zu K\"{o}ln,
Physikalisches Institut, Z\"{u}lpicher Strasse 77, Cologne,
D-50937, Germany}}

\centerline{\footnotesize $^{d}${\it Max Planck Institut f\"{u}r
Radioastronomie, Auf dem H\"{u}gel 69, Bonn, D-53121, Germany}}

\vskip 0.9 truecm

\noindent {\bf Abstract}

\bigskip

The $v$~=~1--0~R(5) line of HD at 2.46~$\mu$m has been detected
at the position of brightest line emission of shocked H$_{2}$ in the Orion
Molecular Cloud. The flux in this HD line, when compared to that of the
previously detected HD 0--0 R(5) line at 19.43~$\mu$m, suggests that, like
the $v$=1 levels of H$_{2}$, the $v$=1 levels of HD are populated in LTE,
despite their much higher rates of spontaneous emission compared to
H$_{2}$. The higher than expected population of vibrationally excited HD
may be due to chemical coupling of HD to H$_{2}$ via the reactive
collisions HD~+~H~$\leftrightarrow$~H$_{2}$~+~D in the shocked gas. The
deuterium abundance implied by the strengths of these lines relative to
those of H$_{2}$ is (5.1$\pm$1.9)~$\times$~10$^{-6}$.

\vskip 0.9 truecm

\noindent {\footnotesize{$^{*}$Corresponding author. Gemini Observatory,
Hilo, HI 96720, USA. Tel. +1-808-974-2519; fax +1-808-935-9650}}

\noindent {\footnotesize{\it Email address:} tgeballe@gemini.edu;
fax: 1-808-935-9650 (T. R. Geballe)}

\vfill\eject

\noindent {\bf 1. Introduction}

\bigskip

Until recently, observations of the deuterium abundance in the
interstellar medium have been restricted almost entirely to ultraviolet
and millimeter wavebands. Both wavebands have limitations: UV observations
probe only regions of low extinction; millimeter measurements are
uncertain because of the inaccurately known degree of chemical
fractionation, which favors formation of deuterated species at low
temperatures. In addition, analyses of measurements in both bands are
plagued by uncertainties in the corrections for line saturation (in the
non-deuterated species).

Recently, two pure rotational lines of HD were detected by the
spectrometers on the Infrared Space Observatory (ISO), toward the Orion
star forming region OMC-1, offering new opportunities for determining
[D]/[H]. Wright et al. (1999) used the Long Wavelength Spectrometer (LWS)
to detect the pure rotational R(0) $J$=1--0 transition at 112~$\mu$m
toward the Orion Bar.  Bertoldi et al. (1999) detected the 19.43~$\mu$m
pure rotational R(5) ($v$=0, $J$=6--5) line at H$_{2}$ Peak~1, the
brightest position of shocked line emission in the OMC-1 outflow, with the
Short Wavelength Spectrometer (SWS).  Both measurements have led to new
estimates of [D]/[H] in Orion, Wright et al. finding
($1.0\pm0.3)~\times~10^{-5}$ and Bertoldi et al. obtaining
($7.6\pm2.9)~\times~10^{-6}$.

Both of these estimates have large uncertainties, stemming only in part
from the low signal-to-noise ratios of the line detections. One source of
uncertainty is the value of [HD]/[H] in the warm partially dissociated gas
produced by C-shocks, where the endothermic reaction,
HD~+~H~+~418K~$\rightarrow$~H$_{2}$~+~D, selectively depletes HD. Bertoldi
et al. (1999) estimate that the depletion factor at Peak~1 is 1.67 for gas
in the range of temperatures where most of the HD $v=0$ R(5) line emission
occurs.  The second is the lack of information about the populations of
other energy levels. This is a serious problem at Peak~1, because (1) the
post-shock gas has a wide range of temperatures, and (2) HD radiatively
relaxes much more rapidly than H$_{2}$, so that its $v$=0 $J$=6 level,
which is 2,636~K above ground, might not be in LTE, unlike H$_{2}$ lines
of similar excitation.

The value of [D]/[H] given by Bertoldi et al. (1999) includes both the
correction for depletion and an adjustment factor of 1.5, their estimate
of the factor by which the population of $v$=0, $J$=6 falls below that for
LTE. This factor, which is highly uncertain, could be tested if other
transitions of HD were detected. One possible source of additional
information is the fundamental vibration-rotation band of HD, which should
be excited into emission by collisions in the post-shock gas. The
fundamental band occurs at the long wavelength edge of the 2~$\mu$m window
and short wavelength edge of the 3~$\mu$m window, where telluric
absorption, mostly by H$_{2}$O, is severe. However, the $v$=1-0~R(5) line
of HD at 2.459~$\mu$m is accessible from a dry, high altitude site such
as Mauna Kea. The upper level of this line is 7,747~K above ground
(Herzberg 1950; Essenwanger \& Gush 1984), and hence the intensity ratio
of it to the 0-0 R(5) line observed by ISO can provide valuable
information concerning the excitation of HD.  The 1-0~Q(5) line of
H$_{2}$, with upper level energy 8365~K (Dabrowski \& Herzberg 1984),
is nearby at 2.455~$\mu$m, so its strength can be easily compared to the
HD line.

\bigskip\bigskip

\noindent {\bf 2. Observations and Results}

\bigskip

A search for the 1-0 R(5) line of HD was conducted on UT 1999 January 20
at the United Kingdom Infrared Telescope (UKIRT), using the 1--5~$\mu$m
spectrometer CGS4 (Mountain et al. 1990). CGS4's echelle was used with a
slit of width 0.82~arcsec oriented east-west across OMC-1 Peak 1 for a
total of 40 minutes of exposure time; an equal amount of time was spent
on blank sky 5~arcmin east. The resolution was 16~km~s$^{-1}$. The
spectrum of the star HR~2007, measured with the same instrumental
configuration, was used to flux- and wavelength-calibrate (using telluric
absorption lines) the spectrum at Peak~1.

The resulting spectrum of a 0.82$\times$7.28~arcsec$^{2}$ strip at Peak 1
is shown in Fig.~1. The spectrum is dominated by the 1-0 Q(5) line of
H$_{2}$, but the 1-0 R(5) line of HD is clearly detected at about a
thousandth the strength of the H$_{2}$ line, roughly as predicted by
Hartquist et al. (1982). The fluxes detected in the above aperture were
$(1.06\pm0.21)\times10^{-15}~{\rm W~m}^{-2}$ and
$(1.2\pm0.3)\times10^{-18} {\rm W~m}^{-2}$, respectively; uncertainty in
the flux calibration dominates the uncertainty in the H$_{2}$ line flux
and is about half the uncertainty in the HD line flux.  The HD line is at
the predicted wavelength to within 0.0001~$\mu$m. Its intensity
distribution, which is similar to that of the Q(5) line along the slit,
solidifies the identification. The profiles of the two lines do not
appear identical, but are the same to within the noise.

\bigskip\bigskip

\noindent {\bf 3. Analysis}

\bigskip

\noindent {\it 3.1 Beam dilution}

\bigskip

To determine the excitation parameters of the HD at Peak 1, it is
necessary to compare the two line fluxes. The 0-0 R(5) line was measured
by the SWS in an aperture of 380~arcsec$^{2}$, centered on Peak~1, whereas
the 1-0 R(5) line was observed at UKIRT through a 6~arcsec$^{2}$ aperture
on the brightest part of Peak~1. From the H$_{2}$ Q(5) line, which was
observed both by the SWS in a large aperture (Rosenthal, Bertoldi, \&
Drapatz 2000) and UKIRT in the small aperture, we conclude that the
observed HD 1-0 R(5) surface brightness must be decreased by a factor of
(2.1$\pm$0.4) to properly compare it with the average surface brightness
of the pure rotational HD line measured in the larger aperture.

\bigskip

\noindent {\it 3.2 Column densities}

\bigskip

The surface brightness of an optically thin vibration-rotation line can
be converted into a column density via
N(v,J)~=~(4$\pi$/hc)~(I$\lambda$/A)10$^{0.4A_\lambda}$), where I is the
observed line intensity, A is the Einstein coefficient, and A$_{\lambda}$
is the extinction, assumed to be 0.78~mag at 2.455~$\mu$m (see Bertoldi
et al. 1999, Rosenthal et al. 2000). Bertoldi et al. found an HD column
density in the $v$=0, $J$=6 level, N(0,6), of
(3.0$\pm$1.1)~$\times$~10$^{14}$~cm$^{-2}$ at Peak 1.  After correcting
for the difference in surface brightnesses as described above, the 1-0
R(5) line yields a column density
N(1,6)~=~(5.1$\pm$1.3)~$\times$~10$^{12}$~cm$^{-2}$ averaged over the
large beam.  This is a factor of 60 lower than the column density in the
$v$=0, $J$=6 level, demonstrating that most of the HD is not
vibrationally excited. However, as shown in Fig.~2, the two populations
match the relative populations of H$_{2}$ levels (Rosenthal et al. 2000;
Bertoldi et al. 1999) of similar energies.  This is somewhat surprising,
as one might have expected that the higher HD levels would have a
sub-thermal excitation, due to their much higher radiative transition
rates (60 times higher when comparing the rate for HD(1,6),
5~$\times$~10$^{-5}$~s$^{-1}$, with that for H$_{2}$(1,4), which has the
same excitation energy). Thus one might have predicted that the
population of the higher energy level of HD would fall far below the
scaled H$_{2}$ curve. We return to the discussion of HD excitation below.

\bigskip\bigskip 

\noindent {\bf 4. Deuterium abundance}

\bigskip

Assuming that the lower energy states of HD are populated in LTE (i.e., in
the same way as the levels of H$_{2}$, as suggested by Fig.~2), and after
correcting for the depletion of HD, we find
[D]/[H]~=~(5.1$\pm$1.9)~$\times$10$^{-6}$, where the uncertainty does not
include the uncertainty in the depletion. This abundance ratio is 1.5
times lower than that given by Bertoldi et al. (1999), who had corrected
for their assumed non-LTE level population by that factor in addition to
correcting for depletion. The abundance is the lowest yet determined for
deuterium. However, within the uncertainties the value is consistent with
the recent value of $7.4_{-1.3}^{+1.9}\times 10^{-6}$ derived by Jenkins
et al. (1999) toward $\delta$~Orionis. We note that Jenkins et al. and
Sonneborn et al. (2000) find that the deuterium abundance varies
significantly along different lines of sight.

\clearpage

\noindent {\bf 5. Excitation of HD}

\bigskip

The excitation of the HD (0,6) and (1,6) levels appears to mimic that of
similarly excited levels of H$_{2}$. If this is generally true for HD,
then there must be a way of maintaining HD level populations in spite of
their radiative transition rates, which are substantially higher than
those of H$_{2}$.  While it is believed that H$_{2}$ levels are in LTE as
a result of standard collisional relaxation with H$_{2}$ and H,
post-shock densities would need to be considerably higher than
10$^{7}$~cm$^{-3}$ in order for simple inelastic collisions to maintain
HD in or near LTE.

We have already noted the exchange reaction,
HD~+~H~$\leftrightarrow$~H$_{2}$~+~D, which tends to deplete HD in the
warm, partially dissociated post-shock gas of OMC-1.  Our analysis of this
reaction, which has been studied previously by Rozenshtein et al. (1985),
Zhang \& Miller (1989), Gray \& Balint-Kurti (1998), and Timmermann
(1996), indicates that at Peak 1 {\it the fraction of vibrationally
excited HD, HD$^*$, is significantly enhanced by this reaction}. Details
of our analysis are provided in Ramsay Howat et al. (2002). To summarize,
we find the following.

\begin{itemize}

\item{H$_{2}$(v=0)~+~D~$\rightarrow$~HD$^*$~+~H is the dominant channel
for "chemically" exciting HD vibrationally, with a rate coefficient of
1.24~$\times$~10$^{-12}$~cm$^{-3}$~s$^{-1}$.  Vibrationally excited
H$_{2}$ has an order of magnitude higher rate coefficient for production
of HD$^*$, but only a very small fraction of H$_{2}$ is vibrationally
excited (see below).}

\item{Formation of HD$^*$ is faster than its relaxation to the ground
vibrational state when
$n$(H$_{2}$)~$>$~4~$\times$~10$^{7}$[$n$(HD)/$n$(D)][$n$(HD$^*$)/$n$(HD)]~cm$^{-3}$.
In this case collisions of HD* with H$_{2}$ and other species will lead to
an LTE population of HD*.  The deuterium fraction, $n$(D)/$n$(HD) is
determined by detailed balance of the HD--H$_{2}$ exchange reaction, and
the above equation simplifies to
$n$(H)~$>$~2~$\times$~10$^{7}$~[$n$(HD$^*$)/$n$(HD)]~e$^{418/T(K)}$.}

\item{The ratio $n$(HD$^*$)/$n$(HD), which is needed to evaluate the
above inequality, varies greatly over the range of post-shock
temperatures where line emission occurs.  For a reasonable assumption of
0.01 for the mean value of this ratio, the formation of HD$^*$ by
reactive collisions is faster than its radiative decay to the ground
vibrational state when $n$(H)~$>$~2~$\times$~10$^{5}$~cm$^{-3}$.}

\end{itemize}

In the post-shock gas $n$(H) is believed to be well in excess of 10$^{5}$
cm$^{-3}$. Thus, we tentatively conclude that at Peak 1 partially
dissociative shocks strongly couple the excitation of HD to H$_{2}$.  
Because of the much larger abundance of H$_{2}$, the HD effectively 
becomes part of the H$_{2}$ level system.

To further test the importance of this excitation mechanism, detections
of additional lines of HD are needed, in order to determine if the
populations of their upper levels also fall on the scaled H$_{2}$ curve
in Fig.~2.

\bigskip\bigskip

\noindent {\bf Acknowledgements}

\bigskip

This program was observed as part of the UKIRT Service Programme. We
thank the staff of UKIRT for its support. The United Kingdom Infrared
Telescope is operated by the Joint Astronomy Centre on behalf of the U.K.
Particle Physics and Astronomy Research Council. We also thank J. Black
and D. Flower for helpful discussions, and C. Wright and an anonymous
referee for their corrections to the manuscript. TRG's research is
supported by the Gemini Observatory, which is opeated by the Association
of Universities for Research in Astronomy, Inc., on behalf of the
international Gemini partnership of Argentina, Australia, Brazil, Canada,
Chile, the United Kingdom, and the United States of America.

\clearpage

\centerline{{\bf References}}

\vskip 0.4 truecm

{\footnotesize

Bertoldi, F., Timmermann, R., Rosenthal, D., Drapatz, S., \&
Wright, C. M. 1999. A\&A, 346, 267. 

Dabrowski, I. \& Herzberg, G. 1984. Can J. Phys, 62, 1639.

Essenwanger, P. \& Gush H.P. 1984, Can. J Phys., 62, 1680. 

Gray, J.K \& Balint-Kurti, G.G. 1998, J Chem. Phys., 108 (3),
950. 

Hartquist, T.W. 1982, in Galactic and Extragalactic Infrared
Spectroscopy, ESA SP 192, 45. 

Herzberg, G. 1950. In Molecular Physics and Molecular Structure. Vol. I
(New York: Van Nostrand Reinhold). 

Jenkins, E.B., et al. 1999, ApJ, 520, 182. 

Mountain, C. M., Robertson, D. J., Lee, T. J., \& Wade, R. 1990,
SPIE, 1235, 25. 

Ramsay Howat, S.K., Timmermann, R., Geballe, T.R., Bertoldi, F.,
\& Mountain, C.M. 2002. ApJ, in press. 

Rosenthal, D., Bertoldi, F., \& Drapatz, S. 2000. A\&A, 356, 705.

Rozenshtein, V.B., Gershenzon, Yu. M., Ivanov, A. V., Il'in, S. D.,
Kucheryavii, S. I., \& Umanskii, S. Ya. 1985. Chem Phys. Lett., 121, 89.

Sonneborn, G., Tripp, T. M., Jenkins, E. B., Sofia, U. J., Vidal
Madjar, A., \& Wozniak, P. R. 2000. ApJ, 545, 277. 

Timmermann, R. 1996, ApJ, 456, 631. 

Wright, C.M., van Dishoeck, E. F., Cox, P., Sidher, S., \&
Kessler, M. F. 1999, ApJ, 515, L29. 

Zhang, J. Z. H., \& Miller, W. H. 1989. J Chem. Phys., 91 (3), 1528.

}

\clearpage

\centerline{{\bf Figure Captions}}

\vskip 0.5truecm

Figure 1 - Spectrum of a 0.82''$\times$7.28'' (NS x EW) area of at the
center of OMC-1 Peak~1 near 2.46~$\mu$m. The H$_{2}$ and HD lines are
indicated. The assumed continuum, used to estimate the flux in the HD
line, is shown by the dashed line. \\

Figure 2 - HD excitation diagram, with column densities plotted versus
energy of upper level. Triangles are pure rotational transitions;
diamonds are vibration-rotation transitions. Most measurements are upper
limits. The curve is the fit to the H$_{2}$ excitation diagram, scaled by
a factor of 6.2~$\times$~10$^{-6}$ to pass through the HD 0-0 R(5) point.

\end {document}
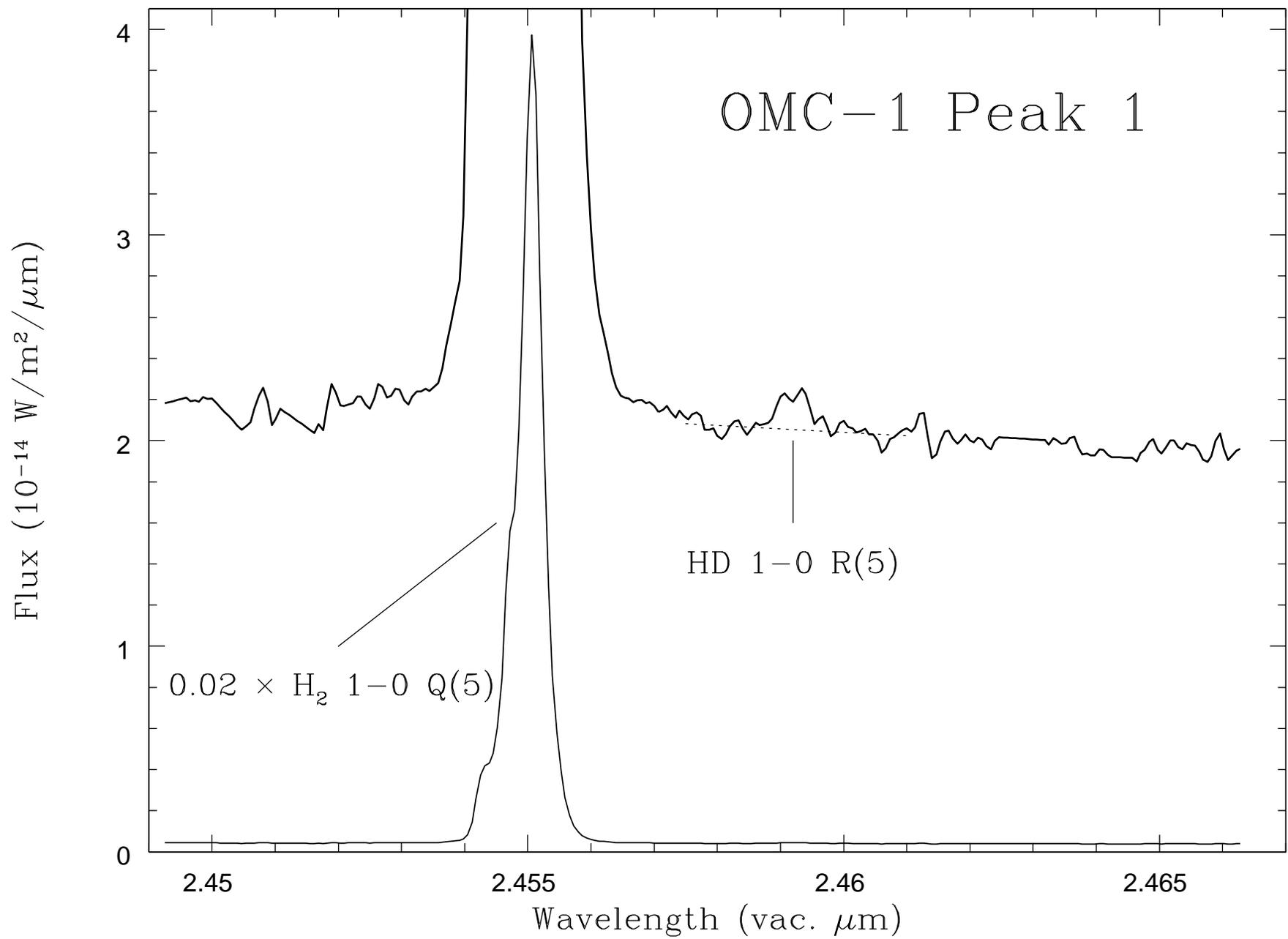
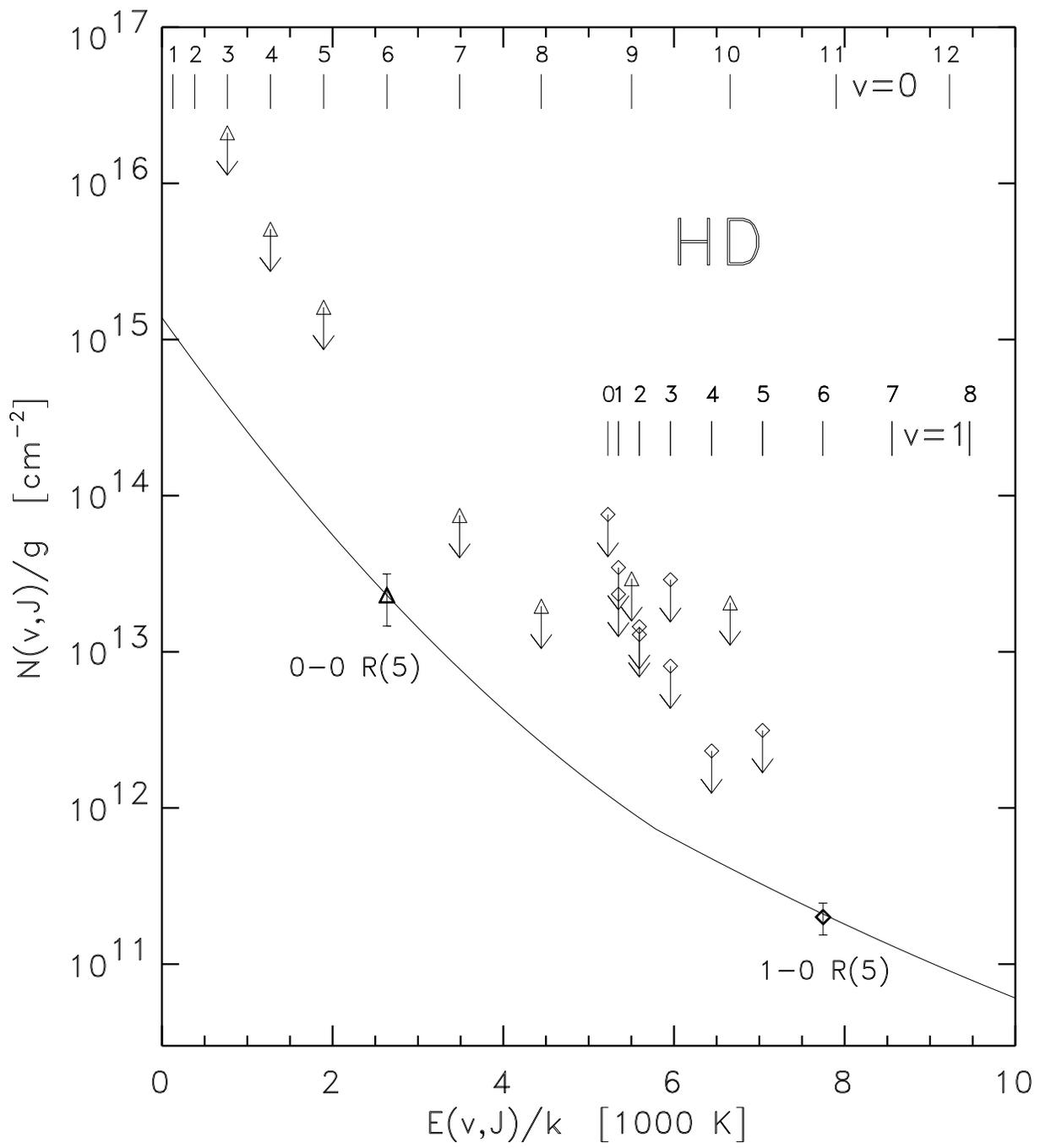
